# Towards a single-junction non-concentrator metal halide perovskite hot carrier solar cell: review of current gaps and opportunities in understanding slow hot carrier cooling

Gabriel J. Man [1,2]

[1] Gabriel Man och Partners AB, Uppsala 752 57, Sweden

[2] Man Labs Technologies LLC, Dallas, Texas 75219, USA

[a] Author to whom correspondence should be addressed:  gman@manlabs.world

**ABSTRACT**

The photovoltaic solar cell is a mature technology, with silicon-based technologies deployed at scale, yet current technologies are limited by the Shockley-Queisser thermodynamic limit, known since the early 1960's.  The single-junction non-concentrator hot carrier solar cell operating at ambient temperature - with its theoretically predicted ultimate power conversion efficiency limit of nearly 70% that is twice the Shockley-Queisser limit and is higher than what can be achieved with even n=6 multijunction solar cells - has remained an elusive yet "hot" research target since the early 1980's.  Metal halide perovskite semiconductors were discovered in the late 1970's and photovoltaic applications have been intensively researched and developed since the early 2010's.  Current technology development of perovskite solar cells is heavily motivated by their expected cheap processing costs relative to other Shockley-Queisser limited technologies.  History has shown that very few absorber materials develop into viable solar cell technologies, and it has been recognized that given the declining costs of silicon-based technologies, a new material must offer potential for both lower cost and higher efficiencies than the Shockley-Queisser limit.  Slow cooling of photocarriers with energy in excess of the band edges (hot carriers), which is the first prerequisite of a solar absorbing material for building a hot carrier solar cell technology, has been reported in perovskites since the 2010's.  *The goal of this review is to illuminate the path towards a single-junction perovskite hot carrier solar cell technology by emphasizing uncertainties in understanding slow hot carrier cooling and recommending approaches to resolve them*.

## I. INTRODUCTION

Since the discovery of the photovoltaic (PV) effect by Becquerel in 1839, the first report of a solid-state PV cell (gold electrodes on selenium) in 1883, and the first report of a silicon-based p-n junction cell with an efficiency of ~6% in 1954, solar cell research and development (R&D) has continually advanced to the present stage where

(predominantly silicon-based) solar cell technologies are deployed at the multi-terawatt scale [1–10]. At present, growing societal awareness of the increasing environmental and human costs related to the combustion of hydrocarbon fuels for generating electricity has given this scientific and technological field new urgency [7]. An important figure of merit (FOM) for terrestrial PV technology is *(light-to-electricity conversion efficiency / true costs ) x lifetime* [2]. Research effort has generally been focused on raising light-to-electricity or power conversion efficiency (PCE), as well as to understand thermodynamic limits for terrestrial solar energy conversion (e.g. Landsberg limit of 93.3%) [4,11,12]. The discovery of new (or application of relatively new) absorber, transport and contact materials has been and continues to be a significant driver for solar cell research activity, though it has been observed that few absorber materials have made the transition into commercially viable solar cell technologies [13].

At present, champion (small area) research-cell PCE's for non-concentrator, single-junction, mature solar cell technologies developed over multiple decades, such as cadmium telluride (CdTe), cadmium indium gallium selenide (CIGS), silicon (Si), and gallium arsenide (GaAs), are 23.1%, 23.6%, 27.4%, and 29.1%, respectively [11,14]. The champion PCE for single-junction metal halide perovskite (MHP) technology, developed intensively since the 2010's, is 27.0%; the theoretical PCE limit of a methylammonium lead iodide (MAPbI$_3$) perovskite solar cell is 30.5% and that of a silicon solar cell is ~29% [15]. Irrespective of the type of solar absorber used, the device architecture used, and the rate of PCE improvement, one could observe that PCE as a function of development time, for most single-junction technologies, is suggestive of a square root function that asymptotically approaches the Shockley-Queisser (SQ) limit of circa 33% PCE predicted for non-concentrator single-junction solar cells [11,16]. This radiative efficiency, or detailed balance, limit considers several intrinsic loss mechanisms (and not extrinsic loss mechanisms due to, for example, material impurities) in a generic PV cell, and was known since 1961. Most semiconducting materials convert sunlight into electricity inefficiently; photon energies in excess of the primary bandgap energy generates nonequilibrium photocarriers (hot carriers) where the excess energy is rapidly dissipated as heat. *The term "hot" refers to the effective temperature used to model the photoexcited carrier density distribution and not to the operating temperature of the solar cell*. A proven approach to surpass the SQ limit is to stack two (tandem) or more (multijunction) solar cells together, where each cell is optimized for absorbing different portions of the solar spectrum; the additional materials and device fabrication complexity, and hence, cost, associated with this approach justifies their use in limited applications.

Other, predominantly research-stage, approaches to surpass the SQ limit focus on processes or materials that enable one of the following two outcomes for photoconversion devices: (i) enhance photovoltage by capturing hot carriers before they cool or (ii) enhance photocurrent by utilizing the excess energy of hot carriers to produce additional carriers [17]. Some refer to these approaches collectively as third-generation PV, where first-generation PV refers to SQ limited silicon wafer-based PV and second-generation PV refers to SQ limited thin film (e.g. CdTe, CIGS) PV [4]. The hot carrier solar cell (HCSC), first reported in 1982, enables the first outcome and essentially combines the functionality of infinite junctions with the simplicity of a single-junction solar cell [18]. This development coincided with general interest in optically-relaxed and optically-excited hot carriers in semiconductors around this time [19]. The HCSC operates by overcoming at least two of the three intrinsic categories of PCE losses that limit conventional solar cells. These categories of losses, visualized in Figure 1, are: (i) non-absorption of sub-bandgap or below-bandgap photons (*pink section*), (ii) photoexcited carrier relaxation to the valence and/or conduction band edge(s) (*green section*) or in other words, cooling of hot carriers into cold carriers, and (iii) other losses (*blue section*). If one simply optimizes for usable electric power as a function of bandgap energy (*black section* in Figure 1), the optimal bandgap energy ranges between ~1.0 to ~1.5 eV. Unsurprisingly, the bandgap energies of Si and GaAs/CdTe, the absorber materials used in the most commercially mature PV technologies at present, are ~1.1 eV and ~1.4 eV, respectively. As shown in Figure 1, the most significant loss categories are (i) and (ii) and a trade-off exists between them: minimization of one maximizes the other. This leads to one of the key postulates in HCSC analysis: that the post-absorption excess energy (photon energy minus bandgap energy) stored in the photoexcited charge carriers is protected from relaxation due to thermal insulation from the environment and is extracted efficiently [18]. From a thermodynamic standpoint, a higher carrier temperature relative to the ambient environment raises the Carnot efficiency. Assuming (ii) can be mitigated, (i) can be minimized by selecting a material with a bandgap < 1.0 eV, leading to a 66% PCE limit that is roughly twice that of the SQ limit.

While evidence for hot electron extraction at a solid-electrolyte heterojunction was obtained shortly after the HCSC analysis was reported, a true, non-concentrator single-junction HCSC operating at ambient temperature has yet to be realized [10,20,21]. Current approaches for developing a working HCSC include varying the absorber material (organic semiconductors, MHP, III-V materials, etc.) [10,21–23], nano-structuring [24–26], using the satellite valleys of the conduction band electronic structure to store the hot carriers [27,28], and introducing quantum wells or using van der Waals heterostructures [29,30]. It was already recognized, ten years ago, that the juxtaposition of declining prices for silicon and the potential of new materials for PV implies that new materials face a difficult cost-benefit challenge to transition to commercial development; they must have potential for both cheap processing costs and SQ limit-breaking performance [17]. Compounding this challenge is the need to meet or exceed the ~25 year lifetimes of commercially available modules. In other words, for a new material to form the basis of a commercial third generation PV technology, it must meet or exceed the lifetime and exceed the PCE of first-generation PV while simultaneously enabling a cost that is lower than that of second-generation PV. As a hybrid combination of a scientific curiosity and a nascent technology, MHPs show potential to meet this challenge. Substantial research effort is ongoing to investigate field-relevant degradation mechanisms and based on available knowledge, the purported material instability of MHPs under the mild conditions to which it is exposed when used in (non-concentrator) solar cells is less a problem than the interfaces [31–34]. Metal halide perovskites can be crystallized from liquid/vapor/solid phase under mild conditions, implying cheap processing costs, and show potential for exceeding the SQ limit, as will be explained shortly.

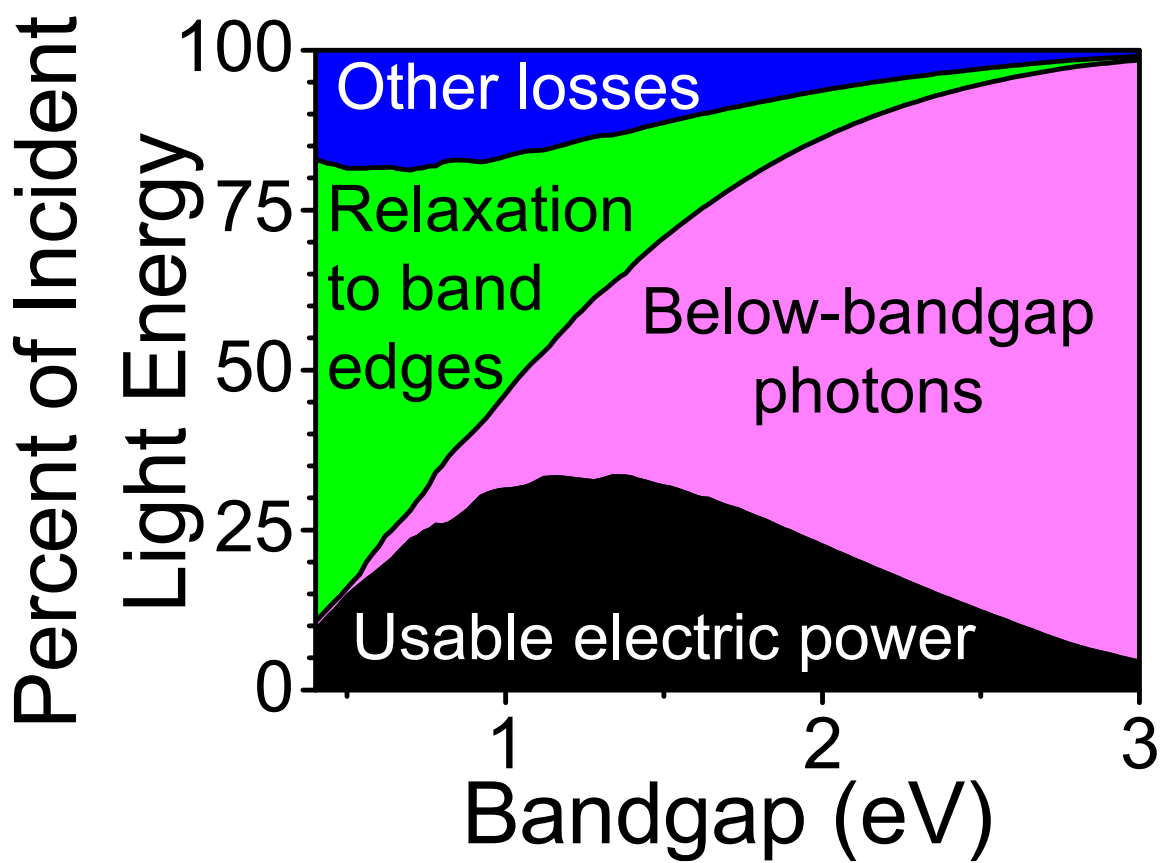

**FIG. 1.** Visualization of the major causes for the Shockley-Queisser limit. The black region represents usable electric power as a function of semiconductor bandgap energy. The green region represents hot carrier energy lost to heat, as a function of semiconductor bandgap energy. The pink region represents light energy not absorbed due to the semiconductor bandgap energy. (From Wikipedia Commons, attributed to Sbyrnes321, used under the Creative Commons CC0 1.0 Universal Public Domain license).

In this work, MHP refers primarily to three-dimensional hybrid organic-inorganic metal halide perovskite materials with the chemical formula $ABX_3$ and cubic-like crystal structures - where $A^+$ is a monovalent organic (e.g. methylammonium (MA)) or inorganic (e.g. caesium (Cs)) cation, $B^{2+}$ is a divalent group IVa cation such as lead II ($Pb^{2+}$) or tin II

($Sn^{2+}$), and $X^-$ is a monovalent halide (e.g. iodide ($I^-$), bromide ($Br^-$)) - that are a sub-family of the general $ABX_3$ perovskite family and have been investigated since 1978 [35–38]. It is speculated that the compounds may have been synthesized at least as early as 1882, before X-ray diffraction for resolving crystal structures became available [39]. The use of MHP as a visible-light sensitizer for titania-based photoelectrochemical cells was first reported in 2009, and the first solid-state MHP solar cell was reported in 2012 [40]. The resurgence of research interest around this time has generated findings that highlight the unique combination of physical and chemical properties, as well as hype [34]. Substantial research interest in MHPs is well summarized in a sentence taken from one report: "lead-halide perovskites are currently the highest-performing solution-processable semiconductors for solar energy conversion, with record efficiencies rapidly approaching that of the Shockley-Queisser limit for single-junction solar cells" [41]. Other types of halide perovskite materials exist, such as ordered perovskites and vacancy ordered perovskites, and many of these materials are currently being researched for solar cell applications [42]. Since 2013, when single-junction MHP solar cells were included in a chart that tracks best research-cell efficiency over time, the PCE has risen from 14.1% to 27.0% [11]. Hybrid silicon/MHP tandem cells were included since 2016, rising in PCE from 23.6% to 34.9%, and the PCE limit of an n=2 tandem solar cell (including MHP/silicon) is predicted to be circa 44% [43,44]. One could interpret this hybrid material tandem direction as an approach for incrementally improving the FOM of silicon solar cell technology beyond what is possible with silicon-only technology.

*However, the most impactful application yet of MHPs in PV may be in a single-junction MHP HCSC.* The discovery of slow hot carrier cooling in MHP's, first reported in 2013, suggests the lifetime of photoexcited hot carriers may be sufficiently long to facilitate hot (and not cold/thermalized) charge extraction, which is a crucial prerequisite for a HCSC [45]. For comparison, hot fluorescence emission with ~$10^2$ ps lifetimes from two organic lead bromide perovskites in monocrystal form was observed, and hot fluorescence emission from bulk GaAs (which forms the basis of the highest PCE single-junction technology to-date) was observed with a time constant of ~$10^1$ ps [29,46]. This suggests the most promising material platform to build a HCSC technology with is MHP, at present. The potential of MHP for HCSCs was quickly recognized shortly after the discovery of slow hot carrier cooling [47–49]. Evidence for hot carrier extraction in a single-junction MHP solar cell was recently reported, albeit with solar concentration and in a device architecture that is not designed for hot carrier extraction [21]. However, uncertainties exist both in our conceptual and quantitative understanding of microscopic hot carrier scattering mechanisms in MHPs, as will be explained later, impeding our ability to design and implement a functional MHP hot carrier solar cell. Accurately understanding the microscopic mechanism(s) of slow hot carrier thermalization determines whether a solar absorbing material can be used in its bulk form, or some form of nano-structuring (e.g. quantum wells, nanocrystals, etc.) is needed. Understanding hot carrier transport (e.g. ballistic, diffusive) and knowing whether hot electron/hole bands exist determines extraction contact design. Knowing how excess photocarrier energy is partitioned between electrons and holes determines whether to optimize the solar cell device architecture for hot electron extraction, hot hole extraction, or both.

The goal of this review is to illuminate a path forward towards *non-concentrator single-junction* MHP HCSC technology operating at *ambient conditions*; this technological form is expected to yield the best possible FOM for MHP (and possibly all existing technologies) since solar concentration adds complexity and cost and single-junction cells are the simplest device structure that can be fabricated. This is achieved by (i) *highlighting where gaps in fundamental understanding currently exist*, (ii) *providing historical examples of approaches, if available, where relevant understanding was gained on other materials such as GaAs, and (iii) providing MHP-specific recommendations to resolve uncertainties*. More so than its potential for realizing lower-cost Shockley-Queisser limited technology, the potential of metal halide perovskites for realizing Ross-Nozik limited technology could motivate continued effort to resolve current limitations in fundamental understanding and reach technology maturation.

## II. CURRENT UNDERSTANDING OF HOT CARRIER EQUILIBRATION AND THERMALIZATION IN METAL HALIDE PEROVSKITES

The conversion of coherent, typically monochromatic laser-induced ultrashort pulsed photoexcitations in a generic semiconductor into cooled free carriers is typically understood as a sequence of four temporally overlapping regimes: (a) coherent regime, (b) non-thermal regime, (c) hot-carrier regime, and (d) isothermal regime [50]. For a generic semiconductor, the coherent regime typically spans ~<200 fs post-photoexcitation and features momentum, carrier-carrier, intervalley and hole-optical-phonon scattering. The non-thermal regime typically spans ~< 2 ps and features electron-hole, electron-optical-phonon, intersubband and intervalley scattering, along with carrier capture in quantum wells. The hot-excitation regime typically spans ~1 to 100 ps; microscopic processes include hot-carrier-phonon interactions, decay of optical phonons, carrier-acoustic-phonon scattering, and intersubband scattering. The isothermal regime, which typically extends beyond 100 ps, involves carrier recombination; this is the regime in which non-hot-carrier solar cells operate. Slight variations of this general picture exist. Reference [4] presents a sequence of eight steps: (1) thermal equilibrium before photoexcitation, (2) coherent stage after pulse, (3) carrier scattering, (4) thermalization of hot carriers, (5) carrier cooling, (6) lattice thermalized carriers, (7) recombination of carriers, and (8) return to equilibrium. Others simply present three steps: (1) photoexcitation, (2) thermalization spanning 10-100 fs, and (3) cooling spanning 1-10 ps [51]. We note, as some others have, that the term thermalization is ambiguous and can mean carrier-carrier thermalization or carrier-lattice thermalization [52]. Henceforth, we use the same definitions as used in the 1982 analysis of the HCSC: (i) *hot non-equilibration* means the carriers are not even equilibrated among themselves, (ii) *hot equilibration* means the excited carriers equilibrate among themselves but not with the surroundings, and (iii) *thermalization* is where the carriers completely equilibrate at the ambient temperature [18]. In other words, hot non-equilibration means the hot carrier distributions are peaked at the excitation energy (in the case of photoexcitation with a monochromatic laser), hot equilibration means the hot carrier distribution is characterized by a carrier temperature distinct from the lattice temperature and electron and hole quasi-Fermi levels, and thermalization is used interchangeably with carrier cooling/relaxation. Since the primary focus is on hot carrier equilibration and thermalization in three-dimensional MHPs, excitonic effects could be disregarded as sub-10 fs temporal resolution pump-probe measurements on MAPbX$_3$ films showed hot exciton to free carrier dissociation within 20 fs after photoexcitation [53].

The sun is a thermal blackbody source with a broad emission spectrum; hence the coherence time of sunlight is short (~1 fs) [54]. The coherent regime (a) is relevant for laboratory experiments utilizing coherent lasers. Hot carriers could in principle be extracted from the non-thermal (b) and/or hot-carrier (c) regimes. Extraction of hot carriers during the non-thermal regime (i.e. the hot electron/hole distribution functions cannot be characterized by a temperature) implies that the quasi-Fermi level splitting, and hence, magnitude of photovoltage, is ill-defined, but a major objective of a hot carrier solar cell is simply to generate photovoltages higher than what can be obtained from a conventional solar cell generating photovoltages from cold carrier distributions.

Linear optical spectroscopy of semiconductors has provided much information on electronic band structure, phonons, plasmons, single-particle spectra and defects, and ultrafast optical spectroscopy has provided much information on non-equilibrium carrier and exciton distributions [50,55]. For example, the earliest observation of radiative recombination from photoexcited hot carriers in gallium arsenide was reported in 1969; the hot carrier cooling rate relative to the radiative recombination rate was sufficiently slow for high-energy tails to be observed in the steady-state photoluminescence spectra [56]. The first evidence for non-thermal distribution functions in photoexcited GaAs was reported in 1985, using pump-probe spectroscopy [57]. The first quantification of hot carrier-energy-loss rates in a semiconductor, conducted with GaAs/ AlGaAs quantum wells, was also reported in 1985 and was enabled by simultaneous measurements of the luminescence spectra and electrical (I-V) characteristics [58]. For context, GaAs solar cells were being developed during this time period, with the first, single-junction, single crystal GaAs solar cell certified at 22.2% PCE in 1977 [11]. To date, the majority of experimental investigations of slow hot carrier thermalization in MHPs utilize optical-pump, optical-probe spectroscopic measurements performed on optical structures (e.g. thin films on optically transparent substrates).  Since similar studies were performed on GaAs over the last few decades, it is worthwhile to study those reports carefully, including and beyond the references cited herein.  Recently, there is a move towards applying transient absorption spectroscopy measurements to MHP device structures [59].

## A. Trends and uncertainties in experimental observations of slow hot carrier thermalization

Hot carriers cool quickly in most semiconductors that otherwise show favourable optoelectronic properties for solar cells, for example in ~0.4 ps in bulk non-nanostructured silicon, 2.7 ps in CIGS, and ~50 fs in bulk GaAs [60–63].  A comparably long hot carrier cooling time of ~0.4 ps was first reported for thin-film $MAPbI_3$ in 2013 using femtosecond transient absorption spectroscopy measurements with 150 fs laser pulses, a photon pump energy of ~3.1 eV ($MAPbI_3$ bandgap energy is ~1.6 eV), and white light continuum probe pulses spanning 1.51 – 2.95 eV [45].  Later studies reported longer cooling time constants.  In 2016, using both time-resolved two-photon photoelectron (tr-2PPE) and photoluminescence (tr-PL) measurements on $MAPbI_3$ films vapor deposited in high vacuum, hot electrons with ~0.25 eV excess energy above the conduction band minimum were found to cool in ~100 ps [64].  In another study, the combination of time-resolved measurements of high energy photoluminescence, which reflects depopulation of the hot states, and transient absorption spectroscopy measurements, which measures the repopulation of the band minima with cold carriers, yielded comparable timescales (~100 ps) from $MAPbI_3$(Cl) films, suggesting hot carrier thermalization occurs without transient trapping of electrons or holes [65].

Experimental trends in MHP composition, such as A-cation substitution, were observed.  Time-resolved photoluminescence measurements of $APbBr_3$ crystals (A = MA, FA, Cs), reported in 2016, yielded hot fluorescence decay constants of 190 ± 20 ps for $FAPbBr_3$, 160 ± 10 ps for $MAPbBr_3$, and no measurable fluorescence for $CsPbBr_3$ (given experimental resolution of 10-20 ps) when pumped with a photon energy of 2.6 eV, roughly 0.3 eV higher than the bandgap energy [46].  A reverse trend, where $CsPbBr_3$ exhibited the longest cooling time, followed by $MAPbBr_3$ then $FAPbBr_3$, was reported in 2018 using ultrafast visible pump-infrared push-infrared spectroscopy measurements on thin films [66].  Similarly, it was reported in 2019, using $APbBr_3$ nanocrystals, femtosecond transient absorption spectroscopy, and the same excitation energy (350 nm or 3.54 eV), $CsPbBr_3$ nanocrystals showed the longest cooling time, followed by $MAPbBr_3$ and $FAPbBr_3$ [67].  More recently in 2024, hot carrier photoluminescence decay time constants of 2.8 ns, 2.0 ns, and 0.6 ns were reported for $FASnI_3$, $MASnI_3$, and $CsSnI_3$ [68].  We note two A-cation trends, one with hot carrier cooling time constants decreasing in this order: FABX$_3$, MABX$_3$, CsBX$_3$, while the other trend is the reverse. It is difficult to reconcile them due to different experimental parameters used, such as sample format (nanocrystals vs. monocrystal vs. films), photoexcitation energy, fluence, etc.

Experimental trends in photoexcited carrier density were observed. In 2015, transient absorption spectroscopy measurements of MAPbI$_3$ revealed that for initial carrier densities above ~6 x 10$^{17}$ cm$^{-3}$, there exists a slowing of the cooling rate consistent with a hot phonon bottleneck effect [69]. Another study published around that time reported a similar threshold, ~5 x 10$^{17}$ cm$^{-3}$, as the critical carrier density for a hot phonon bottleneck-like effect to manifest [70]. An A-cation effect has been observed, where organic A-cation containing MHPs show a ten-fold decrease in thermalization rate compared to an inorganic A-cation containing MHP [71].

To a lesser extent, other time-resolved techniques that utilize non-optical probes have been applied to the study of hot carrier thermalization. In 2016, time-resolved two-photon photoemission spectroscopy (tr-2PPE) measurements on in-vacuum grown thin films of MAPbI$_3$(Cl), which were not exposed to air prior to measurement, revealed that energetic electrons with ~0.25 eV excess energy cooled on the time scale of 100 ps [64]. In 2017, tr-2PPE measurements on in-vacuum cleaved crystals of MAPbI$_3$ were reported, yielding a cooling time of < 1 ps [72]. Large discrepancies are often difficult to reconcile due to a multitude of experimental variations such as sample format and quality, growth conditions, handling history, measurement temperatures, photoexcited carrier densities, etc.

### B. Hot phonon bottleneck, polaronic carrier screening, and Auger heating mechanisms and uncertainties

Slow hot carrier thermalization in MHPs has been attributed to a variety of mechanisms operative at different ranges of photocarrier density, including but not limited to: (i) hot phonon bottleneck effect, (ii) polaron screening effect, (iii) Auger heating effect, (iv) band filling effect, (v) hot-to-cold carrier energy transfer, (vi) diminished electron-phonon coupling, and (vii) an intrinsic phonon bottleneck effect [47,73,74]. As noted by others, mechanisms are operative at different photocarrier densities; at high excitation densities of ≥ 10$^{18}$ cm$^{-3}$, a phonon bottleneck-like effect has been observed, and a screening-like effect has been observed at low excitation densities (10$^{16}$ – 10$^{17}$ cm$^{-3}$) [64]. For the ideal case of ultra-low trap-related (e.g. Shockley-Read-Hall) recombination in MAPbI$_3$ thin film solar cells, steady-state photocarrier concentration ranges between 10$^{15}$ cm$^{-3}$ (operating point) to 10$^{16}$ cm$^{-3}$ (open-circuit), where Auger nonradiative recombination decreases the theoretical external luminescence efficiency to ~95% at open-circuit conditions [15]. Hence, Auger heating and polaron (or, polaron-like but not polaronic) screening effects are operative in MHP solar cells. The photocarrier generation rate reaches $10^{21}$ cm$^{-3}$ s$^{-1}$ [75]. Hence, the phonon bottleneck-like effect could also be operative in the sun-facing regions of the MHP absorber where the rate of light absorption is highest. We briefly review the hot phonon bottleneck, polaron screening, and Auger heating effects here, and refer the reader to the cited review articles for discussions of other mechanisms.

The hot phonon bottleneck mechanism was already studied in the early 1980's, during a time of research interest in the use of high-intensity pulsed lasers to transfer energy to semiconductor surfaces [76]. The critical carrier density, which depends on carrier effective mass and phonon energy, above which the phonon emission frequency falls rapidly (phonon bottleneck) was calculated to be ~6 x $10^{16}$ cm$^{-3}$ for GaAs, a polar direct-gap semiconductor, and ~$10^{21}$ cm$^{-3}$ for silicon, a nonpolar indirect-gap semiconductor. Since the first experimental observations of a phonon bottleneck-like mechanism in MHPs in 2015/16, it has become one of the most investigated mechanisms behind hot carrier thermalization, and ensemble Monte Carlo simulations have revealed the influence of carrier concentration, phonon lifetime, and ratio of phonon absorption and emission scattering rate on cooling time [69,70,73]. The hot phonon bottleneck effect can be modulated; via pulsed terahertz excitation of transverse optical phonons in MAPbI3, time-resolved photoluminescence measurements revealed the cooling time of hot carriers can be prolonged [77]. However, an alternative explanation for the hot phonon bottleneck effect is the Mott transition or polaron density that is analogous to the metal-nonmetal transition observed in doped semiconductors; above a calculated critical carrier density of 3 x $10^{18}$ cm$^{-3}$, the polarons are spatially overlapping and above-bandgap thermal energy is shared between overlapping polaron states and cannot dissipate [52,78]. The Mott polaron density has also been used to explain the microscopic origins of Auger recombination.

The polaron, coined in 1946, was investigated as early as 1933 [79]. Formally, the term polaron refers to the microscopic unit comprising a self-trapped carrier and the associated pattern of displaced atomic equilibrium positions [80]. Polarons whose electronic carriers are not self-trapped are labelled as weak-coupling polarons, and polarons whose carriers are self-trapped are termed strong-coupling polarons. A carrier is self-trapped if its binding energy is greater than the characteristic phonon energy. In 2015, polaronic charge carriers were first hypothesized to exist in MHPs via efforts to rationalize and reconcile various experimental observations, that charge carriers in MHPs are somehow protected from scattering with defects, longitudinal optical (LO) phonons, and other carriers [81]. Calculated polaron coupling constants of 2.4 (electron) and 2.7 (hole), in the range between 1 and 6, suggests intermediate-coupling polarons exist in MHPs [52]. One recent model of a polaron in MHP, proposed as an evolution of the widely discussed Fröhlich polaron model which does not account for efficient charge screening, is a Belgian-waffle shaped ferroelectric polaron that is a large polaron in two dimensions and small polaron in the third dimension [82]. Computational work examining a related MHP compound, $Cs_2AgBiBr_6$, has identified three distinct polaron species: (i) large polaronic helical Bloch points, (ii) small ferrotoroidic polarons, and (iii) periodic twist-density waves [83]. The simultaneous presence of multiple polaron species could be experimentally challenging to validate.

Numerous time-resolved measurements have been attributed to polarons, and attempts are underway to visualize polarons in real-space with techniques such as time-resolved transmission electron microscopy [84]. Transient structural change in the $PbBr_6$ octahedra of $MAPbBr_3$ nanocrystals was inferred from time-resolved X-ray absorption spectroscopy at the lead $L_3$ edge [85]. Radially expanding nanometer-scale strain fields associated with the formation and relaxation of polarons in photoexcited $MAPbBr_3$ single crystal was inferred from femtosecond diffuse X-ray scattering [86]. Photoinduced lattice changes were inferred from time-resolved measurements of $CsPbBr_3$ nanocrystals at the bromine K-edge and lead $L_3$-edge [87]. Photoinduced lattice distortion in $CsPbBr_3$ nanocrystals was inferred from serial femtosecond crystallography [88]. However, evidence contrary to the existence of strong polaronic effects in MHPs exists also. Calculated large electron polaron and bare band electron scattering rates with polar optical phonons are comparable at both low (100 K) and room (300 K) temperature [89]. Evidence from two independent reports, utilizing angle-resolved photoelectron spectroscopy measurements of the valence band dispersion of single crystal $CsPbBr_3$, yielded different conclusions. Measurements of the valence band dispersion in single crystal $CsPbBr_3$ yielded a hole effective mass of $0.26 \pm 0.02$ $m_e$, 50% larger than the DFT calculated bare electron mass $m_0 = 0.17$ $m_e$ [90]. However, another study involving the same type of measurement and single crystals of the same MHP composition questioned the existence of polaronic effects since the experimental effective mass of $0.203 \pm 0.016$ $m_0$ is close in magnitude to the GW calculated $0.226$ $m_0$ [91]. It has been noted that, given 1.2 eV of excess hot electron energy, the calculated per-unit cell specific heat capacity of 1.25 meV $K^{-1}$ at 300 K and the calculated (large) electron polaron radius of ~27 Å that occupies over 300 unit cells, the 3 K increase in temperature is too low to explain experimental observations of much higher carrier temperatures [52].

The Auger effect, an intrinsic nonradiative recombination mechanism, operates whereby the energy released by a recombining electron is immediately absorbed by another electron which then dissipates this energy by emitting phonons [55,92]. Numerous Auger processes, for example phonon-less and phonon-assisted band-band processes and trap-related processes, exist in semiconductors [93]. At photocarrier densities of $10^{18}$ $cm^{-3}$ and higher, the $\sim n^3$ dependence of carrier density dynamics, characteristic of an Auger process, was observed in $MAPbI_3$, $MAPbBr_3$, and $MAPbI_{3-x}Cl_x$ films [94]. As for the microscopic origins, multiple explanations exist. Computed Auger recombination coefficients for $MAPbI_3$ are large relative to other solar absorbers, and is attributed both to a resonance between the bandgap and inter-band transitions to a complex of higher-lying conduction bands, and to the distortions of the $PbI_6$ octahedra [95]. The Mott transition or Mott polaron density, mentioned earlier as an origin of the hot phonon bottleneck-like effect, is another explanation. Dynamic dielectric screening, neglected in earlier computational work, is reported to lower the room temperature Auger coefficient by ~50% in $CsPbI_3$ and $CsSnI_3$ [96]. Uncertainties exist regarding the conceptual and quantitative origins of Auger-type mechanisms.

### C. Alternative non-polaronic mechanism for microscopic carrier screening

We note that for all of the three slow hot carrier thermalization mechanisms discussed earlier (hot phonon bottleneck, polaronic carrier screening, Auger heating), polaron explanations exist. However, evidence against the existence of strong polaronic effects exists also. To complicate matters further, recent work provides an alternative microscopic explanation for polaronic-like but not polaron-originating effects in carrier transport and thermalization, discussed in more detail below [97,98]. The need to understand whether polaronic effects exist in (3D and beyond) MHPs extends beyond the need for accurate understanding of slow hot carrier thermalization; macroscopic quantum optical phenomena such as superfluorescence in (quasi-2D) MHPs has been attributed to spontaneously synchronized polaronic lattice oscillations [99].

In 2015, when the existence of polarons in MHPs was hypothesized, the crystalline structure was viewed as two interpenetrating sublattices, the inorganic $PbX_3^-$ sublattice and the organic (e.g. $MA^+$ or $FA^+$) or inorganic ($Cs^+$) A-cation sublattice [81]. Since it was known that the valence and conduction band states close to the band edges (and thus are relevant for cold carrier transport) are formed by the inorganic $PbX_3^-$ sublattice, the (organic) A-cation sublattice was viewed "…as a medium that modulates the electrostatic landscape experienced by the charge carriers, leading to charge screening and localization" [81]. In 2021, it was reported, via direct X-ray photoexcitation of the nitrogen (1s core level) contained in the $MA^+$ sublattice of $MAPbI_3$, that the conduction band states of the inorganic $PbX_3^-$ sublattice and $MA^+$ A-cation sublattice are electronically coupled, with electron transfer occurring on the timescale of ~1 fs [97]. In 2022, higher-lying states in the conduction band of $APbBr_3$ MHPs (A = FA, MA, Cs) were shown to originate from hybridized (A-cation)-Br states; now the conduction band could be viewed as comprised of two distributions of states offset in electron energy, $PbBr_3^-$ sublattice or bromine 4p σ states at the conduction band edge and higher-lying (A-cation)-Br or bromine 4p π states, and the energetic offset/separation between the peak intensities of these two distributions is the bromine 4p σ-π splitting [98]. The hot phonon bottleneck effect was explained as slowed thermalization of hot electrons (occupying the π states) when a fraction of σ states is occupied, and a positive correlation between hot fluorescence cooling time constant and σ-π splitting was observed.

Based on the two aforementioned reports, it was suggested that (i) the σ states, having Pb-Br character, are delocalized/Bloch-like states whereas the π states, having (A-cation)-Br character, are spatially localized states, (ii) continuously alternating (and possibly lossless) femtosecond charge transfer occurs between the two sublattices, and (iii) a non-thermalized electron undergoing transport and/or thermalization will encounter both σ and π states and exhibit polaronic-like behaviour, alternating between possibly lossless "trapping" and "detrapping" [98]. The alternating femtosecond charge transfer could be relevant for hot electron non-equilibration and equilibration. Through the use of two-dimensional electronic spectroscopy with sub-10 fs temporal resolution, carrier-carrier scattering with time constants < 85 fs was found to be the dominant equilibration mechanism for hot carriers in $MAPbI_3$ [100]. Hot electron occupation of both sublattices, instead of just the $PbX_3^-$ sublattice, should reduce overall Coulombic carrier-carrier scattering via spatial separation. Carrier and trap-resolved photo-Hall measurements of $FAPbI_3$ films, prepared in the same manner as films used in power conversion efficiency competitive solar cells, with trap densities comparable to monocrystals, shows the film is p-type (i.e. holes conduct the majority of the current), and hole mobility, lifetime, and diffusion length exceeds electron mobility, lifetime, and diffusion length across many orders of magnitude of light intensity [101]. The laser light energy is ~2.0 eV, sufficient to generate hot carriers in the ~1.6 eV bandgap $FAPbI_3$. The alternating femtosecond charge transfer in the conduction band could explain the disparities between hole and electron transport; the electrons are subject to alternating femtosecond charge transfer while the holes are not. The distribution of π states could be viewed as a hot electron band, and the operation of a hot carrier solar cell requires hot carrier bands to transport the carriers to the contacts for efficient extraction.

### D. Partition of excess photocarrier energy between hot electrons and hot holes

Excess photocarrier energy is generally partitioned between electrons and holes in such a way that the lighter carrier tends to take a larger share of the excess photon energy [25]. Computed effective masses for holes and electrons in $MAPbI_3$ are somewhat comparable; averaged (along different high-symmetry directions in the Brillouin zone) effective hole and electron masses are 0.25 and 0.19 (GW calculations with spin-orbit coupling), and 0.28 and 0.17 (DFT calculations with spin-orbit coupling) [102]. Details such as the carrier density dependence of effective mass, computed due to band nonparabolicity, exist [103]. Without considering such and other details of the electronic structure, such as higher-lying conduction bands and lower-lying valence bands, one could expect ~60% of the excess photocarrier energy to be distributed amongst hot electrons, and the rest amongst hot holes in $MAPbI_3$. Early observations (2013) of slow hot carrier cooling in $MAPbI_3$ attributed the origin to slow hot hole cooling from a lower-lying valence band (VB2) to the valence band edge (VB1) [45]. In 2015, electron-phonon interaction calculations suggested that long hot hole lifetimes exist in the valence band region 0.6 eV below the valence band edge, and hole relaxation is slowed due to the small valence band density-of-states near the band edge [104]. The negligible effect of A-cation substitution on slow hot hole cooling was predicted also, which contradicts experimental observations of A-cation influence on hot carrier cooling time constants. Alternatively, computational work from 2022 suggests that, in both pristine and point-defective $MAPbI_3$, a hot electron exhibits slower cooling relative to a hot hole since electron cooling proceeds via band-by-band relaxation and hole cooling occurs via direct relaxation [105]. Experimentally, simultaneous photogeneration and relaxation of hot holes and hot electrons in $MAPbI_3$ was observed via time-resolved extreme ultraviolet spectroscopy (via electronic transitions from iodine 4d core level to hybridized iodine 5p valence and conduction band states); when optically pumped with 3.1 eV and with high initial photocarrier densities in the $10^{19}$ to $10^{20}$ $cm^{-3}$ range, ~3/4 of the initial excess energy from the above band gap excitation was transferred to hot holes and ~1/4 of the energy is distributed amongst the hot electrons, and it was found that hot holes cooled faster than hot electrons [106]. Combined ellipsometry measurements and DFT calculations of $MAPbI_3$ shows multiple optical transitions above the fundamental bandgap transition of ~1.6 eV, for example at ~2.5 eV, and given a photoexcitation energy of 3.1 eV, it is unclear how the individual bands where the hot holes and electrons are photogenerated at affects the partition of excess photocarrier energy and overall cooling time constants [107].

### III. APPROACHES FOR RESOLVING UNCERTAINTIES

Uncertainties exist both in our conceptual and quantitative understanding of microscopic hot carrier scattering mechanisms in MHPs, impeding our ability to design and implement a functional MHP hot carrier solar cell. Accurately understanding the microscopic mechanism(s) of slow hot carrier thermalization determines whether a solar absorbing material can be used in its bulk form, or some form of nano-structuring (e.g. quantum wells, nanocrystals, etc.) is needed. For example, $CsPbBr_3$-based multiple quantum wells, with $CsPbBr_3$ layer thicknesses on the order of 1's to 10's of nanometres, separated by organic layers, have been reported to slow hot carrier thermalization [108]. Such structures could be employed if truly needed, though numerous studies discussed earlier suggests a MHP bulk film is adequate; the deposition of a single thin film is more straightforward, and hence, offers potential for lower cost. Understanding hot carrier transport (e.g. ballistic, diffusive) and knowing whether hot electron/hole bands exist determines extraction contact design. Knowing how excess photocarrier energy is partitioned between electrons and holes determines whether to optimize the solar cell device architecture for hot electron extraction, hot hole extraction, or both.

On the computational and theoretical side, general advancements are being made with the calculation of electron-phonon scattering rates using DFT-based methods, coupling time-dependent Boltzmann transport equation solvers (for time evolving out-of-equilibrium carrier distributions) with DFT, etc. [109]. The use of the ensemble Monte Carlo method, solving the Boltzmann transport equation for non-equilibrium carrier distributions, for simulating hot carrier dynamics in MHPs has been attempted [110]. However, quantitative theory and calculations need to be aligned with experiment to truly advance understanding. The following five technical approaches could help resolve experimental uncertainties: (i) using single-crystalline films, (ii) developing and reporting rigorous beam damage checks, (iii) checking the defect-tolerant and self-healing state of the samples being measured on, (iv) performing complementary, time-resolved optical, X-ray, and photoelectron spectroscopy studies with systematic sweeps over a range of photocarrier densities from $10^{15}$ (relevant to solar cell operation) to $10^{19}$ cm$^{-3}$ (high density regime), temperatures, and temporal ranges (e.g. < 1 ps, 1 – 10 ps, 10 – 100 ps with different temporal resolution requirements for each range), and (v) performing the aforementioned time-resolved spectroscopies with both pulsed monochromatic laser energies, swept over a range of photon energies from bandgap energy to near-ultraviolet, and pulsed white light sources. Such systematic studies would aid in confirming/disconfirming microscopic mechanisms, as it is conceivable that multiple mechanisms are operative with overlapping timescales and photocarrier densities. Scientific and technical justifications for the aforementioned five approaches are provided below.

First, measurement artifacts could be minimized through the use of single-crystalline thin films. The global popularity of MHP research is due in part to the availability of an array of low-cost methods for depositing thin films and growing monocrystals, from liquid phase to vapor phase to solid state processes. The nature of these methods also contributes to the lower-cost potential of MHP solar cell technology relative to existing technologies. A consequence of this is that a wide range of sample quality is possible. Some of the earliest MHP solar cell device research, reported in 2013/14, utilized MHP thin films deposited from liquid phase, a method that is still used for fabricating champion efficiency research-grade solar cells [111–113]. While liquid phase deposition is straightforward to perform in the laboratory and was used in a number of studies discussed earlier, substantial complexity exists during crystallization and several processes occur in parallel including solvent evaporation, progressing supersaturation, nucleation, and crystal growth [114]. Lower quality samples could incorporate non-perovskite components/additives, especially at the grain boundaries of polycrystalline films, as well as being more susceptible to beam damage. In principle, single-crystalline samples eliminate the potential for internal contamination. A number of studies discussed earlier have utilized measurements on (and in some cases required) monocrystal samples, where the small crystal size is suitable only for a limited range of experimental techniques. Vapor phase epitaxy of single-crystalline $CsPbBr_3$ thin films was utilized for fabricating trap-free field-effect transistors, with channel lengths ranging from 1-3 mm [115]. Epitaxial growth of several nanometre thick films of $CsPbBr_3$ and $CsSnBr_3$ by molecular beam epitaxy was demonstrated [116]. Such methods enable large-area single-crystalline MHP films with user-defined lateral dimensions and thickness to be deposited, thus enabling a variety of complementary measurements (with different spot size and thickness requirements) to be performed on, in principle, the same (high-quality) sample. Greater flexibility in preparing single-crystalline samples facilitates experiments such as the use of subpicosecond electrical pulses, synchronized to femtosecond optical excitation pulses, to measure the mobility of hot electrons (previously demonstrated on GaAs) [117]. Techniques that require well-ordered crystalline surfaces and feature beam/measurement spot sizes on the order of 100 μm, such as synchrotron-based angle-resolved photoelectron spectroscopy and resonant Auger electron spectroscopy, which have already been successfully performed on millimetre-scale monocrystals, would benefit from a larger number of fresh measurement spots if single-crystalline films were used [91,97]. Hence, large-area single-crystalline thin film is the sample format of choice for fundamental studies.

Second, the development and reporting of protocols for monitoring beam damage is crucial for drawing general conclusions from experimental findings. The susceptibility of MHPs to electron and photon beam damage is known [118–120]. It is estimated that experimental studies published around 2020 and earlier are affected by beam damage to varying degrees [34]. Irrespective of the type of measurement and beam used, one systematic approach reported in some studies involves first performing measurements on high-purity versions of the MHP precursor chemicals (e.g. 99.999% pure lead iodide and >99.99% pure methylammonium iodide, in the case of $MAPbI_3$), at conditions that do not degrade the precursor chemicals themselves (i.e. decomposing lead iodide into metallic lead), to identify characteristic spectral/etc. features, then continuously looping through measurements on the MHP until the spectral/etc. fingerprints of the precursor chemicals are detected [97,98]. This enables the experimenter to estimate the beam dose threshold, above which the measurement becomes unreliable.

Third, awareness and management of the defect-tolerant and self-healing state of the MHP sample improves measurement reliability. Defect tolerance in semiconductors is defined as a material property where the point/structural/etc. defects that exist "…have a very minimal effect on the mobility and lifetime of the electronic charge carriers", and self-healing refers "…to the process of autonomous repair of damage done to a system/material, without need of external factors" [121]. Both properties, speculated to exist in MHPs for years, have been experimentally confirmed [122–125]. The migration of ions is present in MHPs under device-relevant conditions [126,127]. Simultaneous measurements and photodegradation with ~13 keV hard X-rays show the gradual loss of spectral features associated with Pb-Br bonding, due not to the loss of lead but to the loss of bromide, in MAPbBr$_3$ and FAPbBr$_3$, while showing negligible changes to the electronic structure [123]. In addition to providing direct experimental evidence for defect tolerance in the electronic structure, the loss of bromide ions from the probed area leads to an organic APbBr$_3$ that is photodegraded, yet is still a MHP (i.e. has not decomposed into its constituent precursor chemicals). Hence, a standard check for beam damage would be passed, yet the material has been altered. Once the MHP is in the (photo)degraded and defect tolerant state, self-healing presumably occurs autonomously via halide ion migration back to the damaged area, and it is conceivable that halide ion vacancies, accumulation of halide ions, etc. could influence hot carrier dynamics. Simulation with ab initio molecular dynamics shows the interstitial iodide ion (e.g. $I_i^-$) slows down hot electron cooling in MAPbI$_3$ by a factor of 1.5 to 2 [105]. Other computational work claims lead vacancies in MAPbI$_3$ slows hot hole cooling [128]. Systematically monitoring the defect-tolerant and self-healing state of the MHP could be regarded as an extension of checking for beam damage. For MHPs, there exists a minimum of three measurement regimes: (i) pristine, (ii) defect-tolerant and self-healing (reversibly degraded, still relevant for solar cell operation), and (iii) irreversibly degraded.

Fourth, performing complementary time-resolved optical, X-ray, and angle-resolved photoelectron spectroscopy studies in a systematic manner could help to confirm or disconfirm various thermalization mechanisms. The systematic aspect involves sweeping over a range of (i) photocarrier densities from $10^{15}$ (relevant to solar cell operation) to $10^{19}$ cm$^{-3}$ (high density regime), as earlier work points to different mechanisms operative at various photocarrier densities, (ii) temperatures, for exploring effects related to A-cation dynamics or structural phases, and (iii) time scales (e.g. < 1 ps, 1 – 10 ps, 10 – 100 ps with different temporal resolution requirements for each range), as earlier work points to at least two stages of thermalization where the initial cooling is attributed to a mechanism such as LO phonon scattering whereas the latter stage is dominated by the low thermal conductivity of MHPs [52,129]. Femtosecond optical spectroscopy has yielded much information, as the earlier discussion on experimental observations of slow hot carrier cooling shows. Complementing it with time-resolved X-ray and angle-resolved photoelectron spectroscopy enables microscopic mechanisms (potentially operating in parallel) to be confirmed or ruled out. Pump-probe time- and angle-resolved photoelectron spectroscopy (tr-ARPES) is a powerful approach for directly monitoring carrier distribution functions resolved in energy and momentum space, and has been applied to CsPbBr$_3$ (albeit in the high carrier density regime) and other semiconductors relevant for solar absorption such as GaAs [130,131]. Since photoelectron spectroscopy is a technique that is sensitive to 1's to 10's of nanometres of the surface (depending on photoexcitation energy), measurement on electronically passivated surfaces is critical; the presence of surface defects (from processing, environmental contamination, and/or beam damage) leads to nonradiative recombination that could mask the intrinsic hot carrier dynamics. Fortunately, the surfaces of as-crystallized MHPs appear to be electronically self-passivated, as evidenced by low SRV values for $MAPbBr_3$ crystals and trap-free $CsPbBr_3$ field-effect transistors that do not require intentional passivation at the semiconductor/dielectric interface [115,132]. Element-selective and site-specific X-ray spectroscopy is well-suited for multi-element compounds such as MHPs, and time-averaged measurements have already yielded new insight as discussed earlier. Time-resolved X-ray spectroscopy, potentially performed at an X-ray Free Electron Laser facility, offers potential for disentangling attosecond-to-femtosecond-to-picosecond processes involving only the B-X sublattice, or both the A-cation and B-X sublattices (i.e. polarons in the B-X sublattice versus alternating sublattice femtosecond charge transfer). Furthermore, attosecond core-level spectroscopy can be used to differentiate between ultrafast carrier and lattice dynamics [133,134]. The ability to tailor the thicknesses of the deposited large-area single-crystalline film is one crucial enabler for complementary measurements. While single-crystallinity is essential for angle-resolved photoelectron spectroscopy measurements, and optical and X-ray spectroscopy studies benefit primarily from the (expected) higher sample quality of single-crystalline films, optical and X-ray spectroscopy experiments have different film thickness requirements to meet signal-to-noise requirements, depending for example on whether absorption, reflectivity or luminescence is monitored. A second crucial factor is efficient detection. Due to the potential for MHPs to be reversibly/irreversibly altered during measurement as discussed earlier, the use of low beam doses and instrumentation featuring state-of-the-art detection/transmission efficiency to reduce total measurement time are critical. This strongly motivates instrumentation development, for example high sensitivity optical transient absorption spectrometers and higher transmission angle-resolved time-of-flight electron spectrometers [135,136].

Fifth, performing the aforementioned time-resolved spectroscopies with both (i) pulsed monochromatic laser energies, systematically swept over a range of photon energies from bandgap energy to near-ultraviolet, and (ii) pulsed white light sources, would yield both deeper and broader insight. The electronic structure of MHPs, particularly in the conduction band, is complex; one could expect pump-probe experiments performed with different photoexcitation energies to yield different hot carrier relaxation times, mean free paths, and dynamics, which are dependent on different bands and k points in the Brillouin zone. Taking $MAPbI_3$ as an example, optical transient absorption spectroscopy measurements have revealed multiple photoinduced bleach features centered at 1.65 eV, 2.55 eV, and 3.15 eV [137]. Quasiparticle self-consistent GW calculations for $MAPbI_3$, which include splitting of the lead 6p conduction band into sub-bands due to spin-orbit coupling, yield a calculated optical absorption that reflects these features and more [138]. Time-averaged X-ray spectroscopy has revealed two bromine 4p distributions of states in the conduction band of $APbBr_3$ (A = FA, MA, Cs) originating from different bromide chemical bonding, and offset in electron potential energy; it is conceivable that the hot carrier scattering mechanism(s) are different for Pb-Br states versus (A-cation)-Br states [98]. Systematic monochromatic photon energy sweeps, from bandgap energy to near-UV, would help to disentangle the contributions of various sub-bands. Pulsed white light photoexcitation is complementary to pulsed monochromatic laser photoexcitation as solar radiation is not monochromatic, and multiple plate compression transient absorption spectroscopy, with both supercontinuum (~1.4 eV to ~2.5 eV) pump and probe, has been applied to $MAPbI_3$ [139]. Insight from such measurements would be more directly transferable to the operation of a MHP hot carrier solar cell.

## IV. CONCLUSIONS

The cooling of hot carriers generally occurs in the $10^{-2}$ to $10^0$ picosecond timescale in most semiconducting materials used in or of interest for solar energy conversion, but since 2013 has been observed to occur in the $10^{-1}$ to $10^3$ picosecond timescale in metal halide perovskites, depending on material composition, measurement method, sample format, etc. Uncertainties exist both in our conceptual and quantitative understanding of microscopic hot carrier scattering mechanisms in MHPs, impeding our ability to design and implement a functional non-concentrator single-junction MHP hot carrier solar cell. The realization of such a solar cell would conclude a nearly half-century collective scientific journey on a high note, and moreover, is likely a strong requirement for MHPs to form the basis of a commercially viable technology, given current commercial realities such as declining silicon prices and the technological maturity of silicon and thin film technologies.

Femtosecond optical pump-probe spectroscopy and time-resolved photoluminescence, experimental techniques that have historically yielded much information on hot carrier equilibration and cooling in solar cell relevant semiconductors such as gallium arsenide since the 1960's, have contributed much to our preliminary understanding of ultrafast carrier dynamics in MHPs. However, uncertainties exist in experimentally observed trends, and consequently in the theory and computation used to explain experiment. The effect of A-cation substitution on hot carrier cooling time was observed from $APbX_3$ and $ASnX_3$, with A = FA, MA, Cs. We observe that one trend shows cooling time increasing in this order: Cs, MA, FA, whereas another trend shows the reverse. The various experimental parameters (optical pump energy, sample format such as monocrystal or nanocrystal, fluence, etc.) that vary between reports often makes it challenging to reconcile contrasting conclusions. The partition of excess photocarrier energy between hot electrons and hot holes is unclear. Hot hole thermalization was considered to be the primary mechanism in earlier work, challenged by later experimental observations. The electronic structure of MHPs, for example in the conduction band of $APbX_3$, is complex.

A multitude of hot carrier equilibration and thermalization processes have been put forth, with the hot phonon bottleneck, polaron screening, and Auger heating processes being three commonly examined and discussed mechanisms. We observe here that all three mechanisms have polaron explanations. Reports exist also that challenge the existence of polaronic effects in MHPs. We suggest here an alternative carrier screening mechanism, relevant for hot carrier equilibration/thermalization/transport, the *femtosecond alternating metal-halide and (A-cation)-halide sublattice charge transfer mechanism*, that could explain polaronic-like observations without invoking the existence of polarons. This suggestion is rooted in recent work that demonstrates the A-cation does indeed contribute electronic states (primarily to the upper conduction band, in principle accessible to hot carrier-generating optical transitions), and A-cation to metal-halide sublattice charge transfer is possible on the femtosecond timescale, contrary to the view in 2015, when polaronic effects were first hypothesized to exist in MHPs, that the A-cation plays no role in the optoelectronic functionality of MHPs.

Accurately understanding the microscopic mechanism(s) of slow hot carrier thermalization determines whether a solar absorbing material can be used in its bulk form, or some form of nano-structuring (e.g. quantum wells, nanocrystals, etc.) is needed. For example, $CsPbBr_3$-based multiple quantum wells, with $CsPbBr_3$ layer thicknesses on the order of 1's to 10's of nanometres, separated by organic layers, have been reported to slow hot carrier thermalization [108]. Such structures could be employed if truly needed, though numerous studies discussed earlier suggests a MHP bulk film is adequate; the deposition of a single thin film is more straightforward, and hence, offers potential for lower cost. Understanding hot carrier transport (e.g. ballistic, diffusive) and knowing whether hot electron/hole bands exist determines extraction contact design. Knowing how excess photocarrier energy is partitioned between electrons and holes determines whether to optimize the solar cell device architecture for hot electron extraction, hot hole extraction, or both.

To resolve these uncertainties, five technical approaches were put forth. The first is to utilize large-area single-crystalline thin films for complementary, fundamental studies. The second is to develop and report protocols for monitoring beam damage. It has been estimated that studies involving beam-based measurements, published around and before 2020 are affected by beam damage to varying degrees. The third is to monitor the defect-tolerant and self-healing state. These material properties, speculated to exist in MHPs for years, were confirmed around 2020 and onwards. Consequently, for MHPs, there now exists a minimum of three known measurement regimes: (i) pristine, (ii) defect-tolerant and self-healing (reversibly degraded, still relevant for solar cell operation), and (iii) beam damaged (irreversibly degraded). The fourth is to perform complementary time-resolved optical, X-ray, and angle-resolved photoelectron spectroscopy studies in a systematic manner, aiming to confirm or disconfirm various mechanisms.  The systematic aspect includes sweeping over a range of photocarrier densities, from solar cell relevant carrier densities to high excitation densities, sweeping over a range of temperatures, and sweeping over a range of timescales.  Since the 1960's/70's, when ultrafast (laser-based) optical spectroscopies became standard tools for investigating hot carrier dynamics in semiconductors, much development has occurred in complementary experimental techniques such as time- and angle-resolved photoelectron spectroscopy, and attosecond core-level spectroscopy performed with pulsed X-rays generated in the laboratory or at X-ray Free Electron Laser facilities.  Due to the beam-sensitive nature of MHPs, ongoing development of higher sensitivity, higher transmission spectrometers are strongly motivated.  The fifth is to perform the aforementioned time-resolved spectroscopies with pulsed monochromatic laser energies, systematically swept over a range from bandgap energy to near-ultraviolet, for confirming/disconfirming microscopic mechanisms, and pulsed white light sources, for examining hot carrier non-equilibration, equilibration, and thermalization relevant for a hot carrier solar cell.

## ACKNOWLEDGEMENTS

The preliminary ideas for the literature review and technical approaches section evolved from X-ray Free Electron Laser beamtime proposal writing, while GJM was employed as a Zuckerman Postdoctoral Scholar and affiliated with Bar-Ilan University.  The Zuckerman STEM Leadership Program is acknowledged for partial financial support.  Sharon Shwartz, David Cahen, and Hanna-Noa Barad (Bar-Ilan University) are acknowledged for discussions.

## AUTHOR DECLARATIONS

GJM is an owner of Man Labs Technologies LLC, which intends to develop intellectual property around metal halide perovskite solar cell technology.

## DATA AVAILABILITY

Data sharing is not applicable to this article as no new data were created or analyzed in this study.